\documentstyle []{l-aa}

\include{definizioni}
\begin{document}

\thesaurus{ Section 06 (08.07.1, 08.08.1) } %

\title{Theoretical Zero Age Main Sequences revisited}

\author{ V. Castellani \inst{1},  S. Degl'Innocenti \inst{1}$^,$\inst{2}, M. Marconi \inst{3} }

\offprints {S. Degl'Innocenti, Dipartimento di Fisica Universit\`a 
di Pisa, piazza Torricelli 2, 56126 Pisa, Italy, scilla@astr18pi.difi.unipi.it}

\institute{
Dipartimento di Fisica, Universit\'a di Pisa, piazza Torricelli 2,
   I-56126 Pisa, Italy 
\and Istituto Nazionale di Fisica Nucleare, Sezione di Ferrara, via Paradiso 12, I-44100 Ferrara, Italy
\and Osservatorio Astronomico di Capodimonte, via Moiariello 16, I-80131 Napoli, Italy
}

\date{Received 7 December 1998 ; accepted ..........   }

\maketitle

\markboth {Castellani et al.: Theoretical MS revisited} {Castellani et al.: Theoretical MS revisited}

\begin{abstract}

Zero Age Main Sequence (ZAMS) models with updated physical inputs are
presented for selected assumptions about the chemical composition, covering the
ranges 0.6 $<$ M $<$ 1.2 M$_{\odot}$, 0.0001 $<$ Z $<$ 0.04, 0.23 $<$ Y $<$ 0.34. 
The HR diagram location of the ZAMS as a function of Y and Z is discussed both in the
theoretical (logL, logT$_{eff}$) and in the observational (M$_{v}$, B-V) diagrams,
showing that the V magnitude presents an increased dependence on Z to be
taken into account when discussing observational evidences.
Analytical relations quantifying both these dependences are derived.
Implications for the galactic helium to heavier elements
enrichment $\Delta$Y/$\Delta$Z are finally discussed.

\end{abstract}



\section{Introduction}

Long time ago, the observational evidence for MS stars  has been
the very first challenge for the theory of stellar structure 
and the prediction for underluminous metal poor MS 
stars has been among the very first success of the theory. Since
that time, the location in the Color Magnitude (CM) diagram
of the Zero Age Main Sequence (ZAMS)  and its
dependence on the adopted chemical composition keeps being a relevant
ingredient for the investigation of stellar clusters and
in particular for distance determinations through MS fitting.
The issue is now matter of a renewed interest vis-a-vis
the absolute magnitudes made available by the Hipparcos
satellite for a large amount of stars.\\

In this context, theoretical predictions concerning the ZAMS are 
also connected with the still open problem
of the ratio $\Delta$Y/$\Delta$Z marking the enrichment of interstellar medium
during the nuclear evolution of galactic matter.
From an observational point of view, for any given range of metallicities the
location of the related main sequences depends on the
corresponding variation in Y, which thus governs the observed ZAMS
broadening. In spite of the difficulty of the
procedure, which is affected by uncertainties on cluster reddening, metallicity
and distance modulus, several evaluations of the quoted ratio have been 
provided in last decades, by using suitable  relations
between the main sequence thickness and chemical composition
variations (Faulkner 1967, Perrin et al. 1977, Pagel 1995, Cayrel de
Strobel \& Crifo 1995, Fernandes et al. 1996).
However, one has to notice that the related theoretical scenario 
appears far from being firmly established, and the diffuse belief 
that the effects of Y and Z on the ZAMS location cancel
out for $\Delta$Y/$\Delta$Z$\approx$5$\div$5.5 (see e.g. Fernandes et al. 1996,
Mermilliod et al. 1997) runs against the theoretical evidence 
recently given by
Pagel \& Portinari (1998) for which $\Delta$Y/$\Delta$Z =6 should produce
still a not negligible broadening.\\

Owing to the relevance of this issue, in this paper we will revisit
theoretical predictions about the location of ZAMS models both in the 
theoretical (logL, logT$_{eff}$) and observational (M$_{v}$, B-V) diagrams. 
Taking into account
the increasing amount of observational data, the investigation will
be  extended over a rather large range of both Z
and Y values, covering the  ranges Z=0.0001-0.04 and Y=0.23-0.34.
In Sect. 2 we present our models for selected chemical
compositions, whereas in Sect. 3 we derive suitable 
analytical relations,
discussing the implications for the $\Delta$Y/$\Delta$Z ratio.

\section{ZAMS and/or MS models}

As usual, in the following we will use the term  "Zero Age Main Sequence"
(ZAMS) to indicate the HR diagram locus of  stellar models which are
just starting central H burning with the timescale of H consumption
in the stellar interior. More in detail, the term refers to the first
H burning model which has settled in its Main Sequence phase after
having reached the equilibrium of the secondary elements participating in
the various H burning reactions. Accordingly, all these ``Zero Age'' models
have already experienced a phase of nuclear burning, with time scales which
largely depend on the stellar mass tough, in all cases, much shorter than 
the expected central H burning MS phase. In this context, one expects 
that ZAMS stars will evolve increasing their luminosity,
till reaching the exhaustion of central H. However, as discussed by 
Fernandes et al. (1996), for any reasonable assumption about
the stellar ages, one can safely assume that all the stars fainter than
M$_{v}$$\sim$5.5 are practically unaffected by evolution, so that below
this luminosity stars are expected to be in any case close to their
ZAMS location (see also Lebreton et al. 1997 and Pagel \& Portinari 1998).

Bearing in mind such a scenario, we used the FRANEC evolutionary code 
(Straniero \& Chieffi 1991) to compute ZAMS models for selected choices
about the original chemical composition for stellar models covering the mass 
range 0.6 - 1.2 M$_{\odot}$.  The  input physics, but the equation of state (EOS),
is as in Cassisi et al. (1998), who included all the most recent evaluations
of the various physical ingredients given in the literature. The interested 
reader can find in the above quoted paper a complete description of the adopted
physics together with a detailed discussion of  the
influence of the "new" inputs on stellar models. Regarding the EOS, one finds
that the  tabulation by Rogers et al. (1996) used in Cassisi et al. (1998)
does not allow a full coverage of the range of pressures and temperatures required
by our grid of models.

\begin{figure}[htbp]
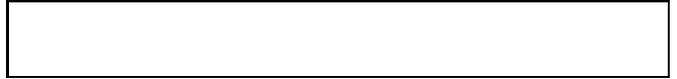
  
\picplace{1cm}
\caption{Comparison among the HR diagram position of ZAMS models
with Z=0.004 by Pols et al. 1998, Fagotto et al. 1994 and by the
present work. Helium abundance as labeled. 
}
\label{conf}
\end{figure}

To overcome this problem,  we adopted the extended EOS tabulation 
given by Straniero (1988) on the basis of the free-energy
minimization method, which takes also into account electron degeneracy and
Coulomb corrections. In the low temperature region we implemented this
EOS with
the Saha equation, which includes the pressure ionization 
contribution,  according
to the method described by Ratcliff (1987). 
Comparison with MS models computed with OPAL
EOS (Rogers et al. 1996), as allowed for selected structures,
shows that Straniero's EOS gives slightly cooler models (by about 100 K) with 
quite a  similar dependence on the adopted chemical composition.
Comparison with similar models presented in the literature, 
as given in Fig. \ref{conf}, shows that
at the larger luminosities our results appear in excellent agreement 
with the recent computations by Pols et al. (1998), becoming redder
at the lower luminosities. This is probably due to the different
EOS, since the above quoted authors adopt an improved version
of the Eggleton et al. (EFF, 1973) equation of state
(see Pols et al. (1995) and Christensen-Dalsgaard \& Dappen (1992) for a 
discussion on EFF EOS).

The difference with models by Fagotto et al. (1994) has probably a similar 
origin, though we have not found in the literature a detailed description
of the EOS used by these authors. Since, in principle, our adopted EOS
should be at least as accurate as the EOS adopted in previous
investigations on the matter, data in Fig. \ref{conf}  can be taken as
an evidence  that a precise location of the ZAMS deserves still more work.
However, comparison among the quoted results discloses a rather
good agreement as far as the effects of  chemical composition
are concerned, so that one can be confident that current models can give a
 realistic information about the dependence on the chemical composition.

\begin{figure}[htbp]
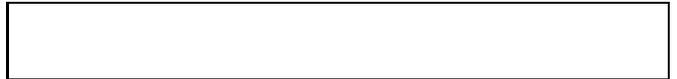
  
\picplace{1cm}
\caption{ZAMS models in the theoretical (left panels) and
observational (right panels) HR diagram for the  labeled assumptions
about the star chemical compositions and for stellar masses ranging
from 0.6 to 1.2 M$_{\odot}$. Data in the left panel assume for the
Sun M$_{bol}$= 4.72.
}
\label{MS}
\end{figure}

Fig. \ref{MS} shows the location,  in both the theoretical (left
panel) and the observational (right panel) HR diagram, of the
present ZAMS, as computed for stellar masses ranging
from 0.6 to 1.2 M$_{\odot}$ and for the labeled values of chemical
composition. Magnitude and colors have been produced 
adopting Kurucz's (1992) model atmospheres which, to our knowledge, 
provide the only available set of models  covering the whole range of metallicities
(up to Z = 0.04) explored in this paper.

As well known, at any fixed effective temperature, both the
theoretical and the observational ZAMS
get fainter as the metallicity decreases or
the helium content increases.  However, Fig. \ref{MS} shows that such 
a behavior appears sensitively enhanced in the observational
plane, due to the dependence of both the color and the bolometric
correction on the star metallicity. Thus the observed MS broadening
cannot properly be discussed on the basis of the behavior of M$_{bol}$
only, 
as sometime given in the literature.

Again in Fig. \ref{MS} one finds that ZAMS run nearly parallel each other
over a rather large portion of the diagrams, allowing the derivation
of analytical relations for the dependence on Y and Z
we will discuss in the next section.

In this context, one has to notice that the HR diagram location of the
models is dependent on the assumption about the efficiency of superadiabatic
convection which affects the external layers of most 
of the models in Fig. \ref{MS}.  To minimize this problem we will restrict 
our investigation (see Sect. 3)
to the lower portion of the main sequence, where superadiabaticity
effects are expected to be negligible (see, e.g., VandenBerg et al. 1983,
Pagel \& Portinari 1998).

\section {The Dependence of ZAMS position on Y and Z} 
Interpolation among the bolometric magnitudes at logT$_{eff}$=3.70
gives:\\

${\rm {M_{bol}}_{(logT_{eff}=3.70)} } = 3.219 + 2.425~{\rm Y} -1.411~{\rm logZ}$
\begin{equation}
 -0.176 {\rm (logZ)}^2
\end{equation} 
which is valid over the whole range of explored compositions with a R.M.S = 0.010
and which can be used to constrain the shift of the ZAMS when varying Y and/or Z.

Note that the  fit of all the data requires a quadratic
dependence on logZ.
 When considering the dependence of the 
visual magnitude
on Y and Z at fixed B-V one cannot find a reference color 
for a suitable coverage of  the ZAMS all over the explored range 
of metallicities (see Fig. \ref{MS}).
Thus we choose two B-V values according to selected ranges of
metallicities, namely : B-V=0.8 for Z $\le$ 0.01 and B-V=1.1 for Z $\ge$ 0.01.  
The two corresponding relations are:\\

${\rm M_v}_{(Z \le 0.01, B-V=0.8)}=-0.184+16.064 Y +(-6.258$
\begin{equation}
 + 15.686~{\rm Y} ) \cdot {\rm logZ} + (-1.433 + 4.551~{\rm Y}) \cdot {\rm (logZ)}^2   
\end{equation}
with a R.M.S.=0.013.\\

${\rm M_v}_{(Z \ge 0.01, B-V=1.1)} = 2.105 + 1.404~{\rm Y} + (-1.861$
\begin{equation}
 - 1.542~{\rm Y}) \cdot {\rm logZ} + (-0.179 - 0.373 {\rm Y}) \cdot ({\rm logZ})^2
\end{equation}
with a R.M.S. = 0.005.\\

However, at M$_{v}$=6 mag, one can find an analytical relation 
connecting the B-V values to Y and Z, all over the explored range of chemical compositions, as given by:\\

${\rm (B-V)_{(M_v=6)} } = 1.663 - 0.473~{\rm Y} + 0.467~{\rm logZ}$
\begin{equation}
 + 0.055 {\rm (logZ)}^2 
\end{equation}
with a R.M.S.= 0.004.\\

\begin{figure}[htbp]
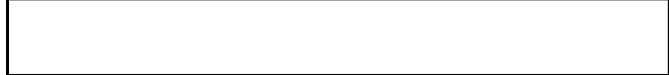
  
\picplace{1cm}
\caption{
Left panel: the relation between Y and Z in the assumption
of $\delta$M$_{bol}$=0 with respect to the solar (Z=0.02 Y=0.27)
value at fixed temperature (logT$_{eff}$=3.70) (see text). The dashed
line indicates $\Delta$Y/$\Delta$Z= 5 relation. Right panel: 
as in the left panel but for $\delta$(B-V)=0 at M$_{v}$=6 mag.  
}
\label{DYDZ}
\end{figure}

Now we are in the position of discussing the predicted effects of
the various correlations between Z and Y. 
Let us first take as ``reference value'' the bolometric magnitude for
the solar ZAMS (Z=0.02, Y=0.27) at fixed temperature
(logT$_{eff}$=3.70) to investigate which variation of Y could be able
to balance a variation in metallicity. Note that this is just a theoretical experiment,
since we know that metallicity does cause a spread in the ZAMS location (see
e.g. again Pagel \& Portinari 1998).\\

Fig. \ref{DYDZ} (left panel) shows, for each given assumption
on Z, the values of Y for which one has the same bolometric magnitude 
(at logT$_{eff}$=3.70) as the Z=0.02 Y=0.27 ZAMS, that is the value of Y for
which one would expect no spread of the MS due to chemical composition.
The dashed line indicates the $\Delta$Y/$\Delta$Z
= 5 relation. One may notice that in this case present theoretical
predictions are in reasonable agreement with the slope
$\Delta$Y/$\Delta$Z$\approx$5, as often referred in the literature,  but
only in a restricted metallicity range around Z=0.02. However, the
right panel of Fig. \ref{DYDZ} shows the relation between Y and Z 
needed to keep unchanged the ZAMS color at M$_{v}$=6.  Now one finds that 
a much larger value  of $\Delta$Y/$\Delta$Z  (of the order of 7 around the 
solar metallicity) would be required in order to avoid a broadening of MS with Z,
showing the additional contribution of model atmospheres to the spread
of visual magnitudes. Such a result appears in agreement with the already quoted 
Pagel \& Portinari (1998) finding, for which $\Delta$Y/$\Delta$Z = 6 is still not sufficient
for avoiding the spread.

\begin{figure}[htbp]
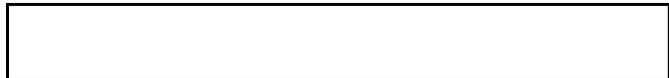
  
\picplace{1cm}
\caption{Hipparcos data for stars with Z$\leq$0.02 in the M$_{v}$-logT$_{eff}$
plane (from Pagel \& Portinari 1998) compared with our theoretical ZAMS 
with $\Delta Y$/$\Delta Z$ = 2, see text. The accuracy on the temperature
observational data is assumed to be $\pm$ 50 K (see Pagel \& Portinari
1998).  The continuous lines are our theoretical ZAMS for Z=0.02, Y=0.27
(upper line) and Z=0.0001, Y=0.23 (lower line), whereas the dashed lines give
the ZAMS for Z=0.01 Y=0.25 (upper line) and Z=0.001, Y=0.232 (lower line). Black dots
show the predicted magnitudes at logT$_{eff}$= 3.70 for the above quoted composition 
and (at the larger luminosities) for the two suprasolar compositions Z = 0.03, 
Y=0.29;  Z=0.04, Y= 0.31. }
\label{disco}
\end{figure}

As a final point, one may explore the rather popular assumption 
that the galactic enrichment follows 
a linear relation between Y and Z, adopting  Z=0.0001 Y=0.23 for 
metal poor stars and Z=0.02, Y=0.27 for the sun, so that
$\Delta Y / \Delta Z \sim 2$.  
In Fig. \ref{disco} we make use of our predictions under the above quoted assumption
to repeat the analysis given by Pagel and Portinari (1998)
for Hipparcos MS nearby stars. One finds that the predicted spread 
appear in fair agreement with observations.
This is not a surprising result, since  Pagel
and Portinari already  found a reasonable agreement
for the rather large range of values  $\Delta Y / \Delta Z$ = 3 $\pm $2.
One finds also several concordances concerning the relation between the visual magnitude
(at fixed temperature) and metallicity.
However, it is difficult to go deeper in such a discussion, owing to the rather large
uncertainties still existing in the observational sample.

\section{Conclusions}

We have investigated the theoretical scenario for ZAMS stars 
with different chemical compositions, providing analytical relations
connecting the MS location  to Y and Z. We show that the assumption of no
MS broadening for $\Delta$Y/$\Delta$Z =5 is supported by theory
only for the behavior of bolometric magnitudes. 
On the contrary, we find that the no-broadening condition
can be reached in the observational
M$_{v}$,B-V plane only if $\Delta$Y/$\Delta$Z$\simeq$7.

\section{Acknowledgments}

It is a pleasure to thank Joao Fernandes for useful discussions during the X Canary Islands
Winter School of Astrophysics and Onno Pols for kindly providing us with the results of his 
computations. This work was partially supported by CNAA through a postdoc research grant to
M. Marconi.

\end{document}